\documentclass[alpha-refs]{wiley-article}

\usepackage{graphicx}
\usepackage{xcolor}
\usepackage[space]{grffile}
\usepackage{latexsym}
\usepackage{textcomp}
\usepackage{longtable}
\usepackage{tabulary}
\usepackage{booktabs,array,multirow}
\usepackage{amsfonts,amsmath,amssymb}
\usepackage{natbib}
\usepackage{url}
\usepackage{hyperref}
\usepackage{float}
\hypersetup{colorlinks=false,pdfborder={0 0 0}}
\usepackage{etoolbox}
\newif\iflatexml\latexmlfalse

\AtBeginDocument{\DeclareGraphicsExtensions{.pdf,.PDF,.eps,.EPS,.png,.PNG,.tif,.TIF,.jpg,.JPG,.jpeg,.JPEG}}

\usepackage[utf8]{inputenc}
\usepackage[english]{babel}

\usepackage{siunitx}

\usepackage{subcaption}
\usepackage{color, colortbl}
\definecolor{Gray}{gray}{0.9}
\usepackage{amssymb}
\usepackage{amsmath}
\usepackage{caption}
\iflatexml

\else
\abbrevs{ABL, KIAB, LHR, LTM \\ \\
Currently at: $^{2}$ Council on Energy, Environment and Water, 4, ISID Campus, Vasant Kunj, New Delhi 110070 India} 

\corraddress{K R Sreenivas, Engineering Mechanics Unit, Jawaharlal Nehru Centre for Advanced Scientific Research, Jakkur PO, Bengaluru 560064, India}
\corremail{krs@jncasr.ac.in}
\presentadd{ $^1$   
Engineering Mechanics Unit, Jawaharlal Nehru Centre for Advanced Scientific Research, Jakkur PO, Bengaluru 560064, India}  

\fundinginfo{a. Technical Research Centre, Jawaharlal Nehru Centre for Advanced Scientific Research (JNCASR), Bengaluru 560 064, India,  \\ 
b. Bengaluru International Airport Limited, Bengaluru 560 300, India  AND \\
c. National Supercomputing Mission, JNCASR, Bengaluru 560 064, India}
\fi

\papertype{Research Article}

\title{Investigation of the Thermal Structure in the Atmospheric Boundary Layer During Evening Transition and the Impact of Aerosols on Radiative Cooling}

\author[1]{Suryadev Pratap Singh}
\author[1, 2]{Mohammad Rafiuddin}
\author[1]{Subham Banerjee}
\author[1]{Sreenivas K R}

\affil[1]{Engineering Mechanics Unit, Jawaharlal Nehru Centre for Advanced Scientific Research, Jakkur, Bengaluru 560064, India}

\runningauthor{SINGH ET AL.} 

\begin{document}

\maketitle
\selectlanguage{english}
\newpage
%\clearpage
\begin{abstract}
The evening transition is crucial in various phenomena including boundary layer stability, temperature inversion, radiation fog, vertical mixing, and pollution dispersion. We have explored this transition using data from eighty days of observations across two fog seasons at the Kempegowda International Airport, Bengaluru (KIAB). Through field experiments and simulations integrating aerosol interaction in a radiation-conduction model, we elucidate the impact of aerosols on longwave cooling of the Atmospheric Boundary Layer (ABL).

Field observations indicate that under calm and clear-sky conditions, the evening transition typically results in a distinct vertical thermal structure called the Lifted Temperature Minimum (LTM). We observe that the prevailing profile near the surface post-sunset is the LTM-profile. Additionally, the occurrence of LTM is observed to increase with decreases in downward and upward longwave flux, soil sensible heat flux, wind speed, and turbulent kinetic energy measured at two meters above ground level (AGL). In such scenarios, the intensity of LTM-profiles is primarily governed by aerosol-induced longwave heating rate (LHR) within the surface layer. Furthermore, the presence of clouds leads to increased downward flux, causing the disappearance of LTM, whereas shallow fog can enhance LTM intensity, as observed in both field observations and simulations.

Usually, prevailing radiation models underestimate aerosol-induced longwave heating rate (LHR) by an order, compared to actual field observations. We attribute this difference to aerosol-induced radiation divergence. We show that impact of aerosol-induced LHR extends hundreds of meters into the inversion layer, affecting temperature profiles and potentially influencing processes such as fog formation. As the fog layer develops, LHR strengthens at its upper boundary, however, we highlight the difficulty in detecting this cooling using remote instruments such as microwave radiometer.

\textbf{Keywords} --- Aerosols, radiation divergence, fog, longwave cooling, lifted temperature minimum
\end{abstract}
%%%%%%%%%%%%%%%%%%%%%%%%%%%%%%%%%%%%%
\section{Introduction}
The evening transition in the Atmospheric Boundary Layer (ABL) holds practical significance, as highlighted in various studies \citep{grant1997observational, fernando2013phoenix, angevine2020transition}. Post-sunset, this transition leads to the formation of the stable Nocturnal Boundary Layer (NBL), influencing several meteorological phenomena including inversion layer growth, fog occurrence, and the impact of stable layers on vertical mixing and pollution dispersion. The cooling rate and moisture accumulation within a few meters above ground level in the NBL are crucial for determining the onset, progression, and dissipation of radiation fog \cite{gultepe2008fog}. Additionally, cooling rates affect the occurrence and strength of temperature inversions, complicating pollution dispersal \cite{hou2016long}. Surface-heat flux, a key factor in dew formation at night, is regulated by near-surface temperature and relative humidity \cite{monteith1957dew}. Therefore, comprehending and analyzing the evening transition and its detailed characteristics through vertical temperature and radiative cooling rate profiles holds practical importance. 

Atmospheric turbulence subsides around sunset and a stable inversion layer forms. According to conventional explanations found in textbooks \cite{stull1988introduction}, the process starts with radiative cooling of the ground and subsequent cooling of the air layers above it, with lowest temperature occurring on the ground. However, an intriguing observation by \citet{ramdas1932vertical} under calm and clear sky conditions is that the ground does not attain the lowest temperature locally. Instead, a local minimum temperature, known as the Lifted Temperature Minimum (LTM), appears a few decimeters above the ground surface. The height at which this minimum occurs is referred to as LTM height, and the difference between the surface temperature and the LTM is termed LTM intensity. 

Despite the robust observations of the LTM in numerous field experiments worldwide, it took considerable time to provide an explanation for its occurrence. \citet{ramdas1932vertical} and \citet{ramanathan1935derivation} initially hypothesized the role of radiation in LTM formation, but this was met with skepticism due to several reasons: (a) it contradicted the prevailing belief that the ground cools faster than the surrounding air layers after sunset, (b) the challenge of maintaining LTM against convective instability for extended periods, and (c) alternative explanations such as drainage flow or measurement errors \cite{geiger1957climate}. However, over time, \citet{lake1956temperature, ramdas1957nachtliche, oke1970temperature, mukund2010hyper, mukund2014field, blay2015lifted, jensen2016observations} conducted meticulous field experiments on various types of soils in different regions of the world. These experiments confirmed the robust and widespread occurrence of LTM across diverse terrains, including snow, bare soil, grassland \cite{lake1956temperature, oke1970temperature, blay2015lifted}, concrete surfaces \cite{mukund2010hyper, mukund2014field}, and mountainous terrain \cite{jensen2016observations}, with varying LTM intensities. Moreover, they refuted the role of drainage flow in LTM development through precise measurements of local winds. By manipulating surface properties in field experiments, \citet{mukund2014field} demonstrated that LTM intensity is strongly influenced by surface emissivity and thermal properties. Additionally, their controlled laboratory experiments, which avoided drainage flow, conclusively showed that LTM intensity decreases with lower aerosol concentrations. They observed significant radiative cooling in the air layer adjacent to the ground, highlighting the importance of aerosol-induced radiation divergence in LTM development. This phenomenon underscores the complexity of nocturnal temperature profiles and emphasizes the significance of radiative cooling in modeling the nocturnal atmospheric boundary layer. 

Apart from the field experiments, there have been efforts to develop mathematical and numerical models to understand the origin and parametric dependence of the LTM on other factors. For instance, \citet{varghese2003fast} modified the radiation model by \citet{chou1993one}, incorporating energy transfer from radiation, conduction, and forced convection into a mathematical model. However, \citet{edwards2009radiative, ponnulakshmi2012hypercooling, ponnulakshmi2013hypercooling} identified erroneous assumptions in \citet{varghese2003fast}'s treatment of ground reflection, particularly regarding the downward longwave (LW) flux, leading to spurious cooling. Subsequent corrections, as proposed by \citet{ponnulakshmi2012hypercooling, ponnulakshmi2013hypercooling}, failed to replicate an LTM profile. \citet{mukund2010hyper} suggested the inclusion of aerosol-radiation interaction to explain the LTM phenomenon, a proposition supported by conclusive experimental evidence from \citet{mukund2014field}. Building on the findings of \citet{mukund2014field}, we incorporated aerosol-radiation interaction into the corrected model of \citet{edwards2009radiative, ponnulakshmi2012hypercooling, ponnulakshmi2013hypercooling} to simulate the thermal structure in the nocturnal boundary layer. Further details of the models are provided in Section \ref{sec: rad_model}.

Despite the significant role of radiation divergence in the Nocturnal Boundary Layer (NBL) under low-wind conditions \citep{gopalakrishnan1998study, mahrt1985vertical, garratt1981radiative, drue2007characteristics} and its importance in numerical weather prediction (NWP) \citep{steeneveld2006modeling, steeneveld2014current, ha2003radiative}, current radiation parameterizations in NWP models fail to accurately simulate the evolution of radiative fluxes \citep{steeneveld2008evaluation, steeneveld2010observations}. Additionally, radiative cooling rates reported in many field experiments lack robustness due to measurement uncertainties and are comparable to reported longwave (LW) cooling rates \citep{funk1960measured, lieske1967measurements, fuggle1976long, nunez1976long, nkemdirim1978comparison, xing1983study, nkemdirim1988nighttime, stull1988introduction}. However, a set of observations by \cite{hoch2005radiative, hoch2007year, drue2007characteristics, steeneveld2010observations} have conducted careful measurements and reported significant radiative cooling within a few tens of meters near the ground during the night.

In addition to measurements, numerous researchers have endeavored to simulate radiative cooling using various methods \citep{funk1961numerical, zdunkowski1965infrared, garratt1981radiative, estournel1985influence, steeneveld2006modeling, steeneveld2010observations}. However, these modeling approaches have encountered challenges such as limited vertical resolution, and unrealistic assumptions, or adjustments to parameterizations to capture observed radiative cooling \citep{raisanen1996effect, steeneveld2014current}. \citet{ha2003radiative} noted discrepancies between estimated radiative and turbulent flux divergence and observed cooling in the Stable Boundary Layer (SBL). While \citet{steeneveld2006modeling} found reasonable agreement between modeled radiative cooling rates and CASES-99 observations, however, parameterization coefficients for modeling needed to be adjusted for getting agreement. Through meticulous observations, \citet{steeneveld2010observations} and \citet{sun2003heat} reported high radiative cooling rates, particularly during evening transitions and under clear sky conditions. They also highlighted that commonly used longwave radiation models underestimate observed cooling during the evening transition by an order of magnitude, whereas a physical model \citep{coantic1971interaction} showed better agreement without aerosols. However, this physical model's applicability is limited due to assumptions like logarithmic temperature and humidity profiles, stationarity, and other parameterizations based on Monin-Obukhov theory \citep{coantic1971interaction, mahrt2014stably}, rendering it less suitable for the use in NWP models \citep{steeneveld2010observations}.

The observed deficiencies in radiative parameterization have been highlighted in both observational and numerical investigations \citep{wild2001evaluation, rinke2012evaluation, steeneveld2014current}. While \citet{zdunkowski1976one, coantic1971interaction, andre1982nocturnal, mukund2010hyper} have speculated on the potential significance of aerosols in longwave radiative modeling, laboratory experiments by \citet{mukund2014field} demonstrated the necessity of considering aerosols to explain temperature profiles near the ground. Despite the dominance of LW radiation divergence over turbulent flux divergence in low-wind conditions, as observed in our study \citep{gopalakrishnan1998study, mahrt1985vertical, garratt1981radiative, drue2007characteristics}, aerosols have not been incorporated into LW radiation modeling to elucidate observed radiative cooling in the nocturnal boundary layer.

In summary, the preceding discussion underscores the necessity of conducting field experiments alongside numerical simulations, incorporating aerosol-radiation interaction, to elucidate the influence of aerosols on radiative processes in the nocturnal boundary layer. Additionally, it aims to determine the height within the atmosphere, where aerosols impact thermal structure. In this pursuit, we present field observations of evening transitions and nights spanning an extensive eighty-day period across two fog seasons. We integrate aerosol-radiation interaction into the corrected band model \cite{ponnulakshmi2013hypercooling}. Through simulations and field observations, we explore the impact of aerosol-induced cooling on evening transition and LTM in calm and clear sky conditions.

This paper is organized in the following way: Field experiments, instrumentation details, and observational data are presented in Section \ref{section: Field Campaign and data}. The longwave (LW) radiation model and integration of the aerosols model into it are elaborated in Section \ref{sec: rad_model}. LTM observations and analysis, LHR during evening transitions, and the influence of fog/cloud on LTM are presented in Section \ref{Sec: Result_section}. Limitations of current radiation models and observations regarding LTM and LHR have been discussed in Section \ref{sec: discussion}, and last, we conclude this work in Section \ref{sec: conclusion}.

%%%%%%%%%%%%%%%%%%%%%%%%%%%%%%%%%%%%%%%%%%%%%%%%%%%%%%%%%%%%%%%%
\section{Field campaign and observational data} \label{section: Field Campaign and data}

\begin{figure}[H] 
	\centerline{\includegraphics[width=\textwidth]{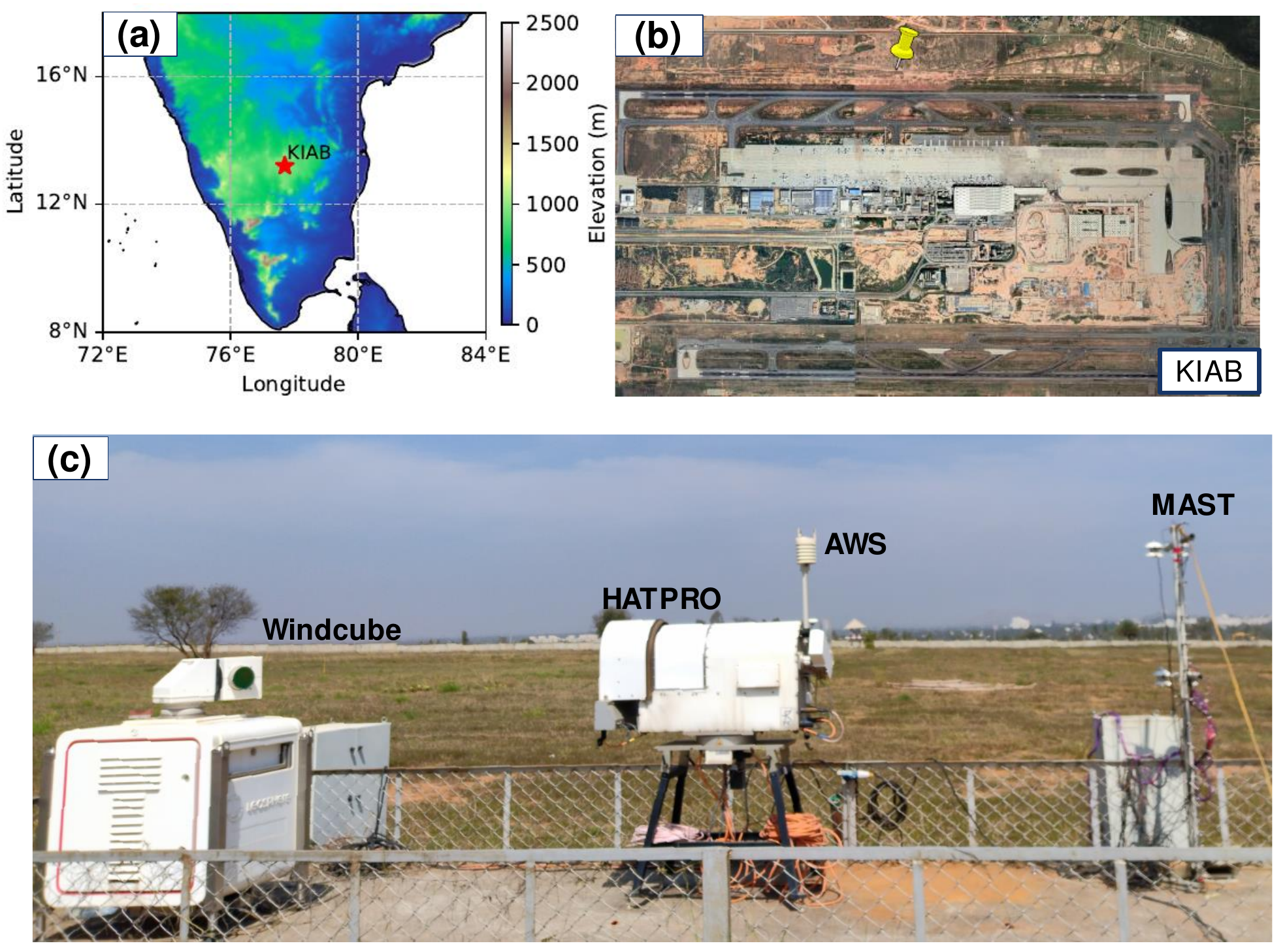}}  
	\caption{Observation site and mounted instruments. (a) Field experiments have been conducted in the airfield of KIAB (Red star) (obtained by $\mathrm{Google-Earth ^{TM}}$). (b) Aerial view of KIAB, observation site (dropped pin), and (c) Instruments installed on the concrete base.}
	\label{fig: observation site}
\end{figure}

The field campaign has been conducted in the airfield of Kempegowda International Airport, Bengaluru (KIAB), located at 13.20$^{\circ}$N, 77.70$^{\circ}$E and $\sim$900 m above mean sea level (Figure \ref{fig: observation site}a). The observation site is $\sim$175 m north of the north-runway (09L/27R; pin dropped in Figure \ref{fig: observation site}b). All instruments (except soil temperature profiler and sensible heat flux sensors) are mounted on a concrete base (9 m $\times$ 3 m) to maintain their alignment and orientation during the campaign, whereas soil sensors are installed into the soil, 0.5 m away from the concrete base.  Although sensors are installed either directly on the concrete base or in close proximity to it, measurements may still differ from those taken on soil. Nevertheless, the thermophysical and radiative properties of soil are comparable to those of concrete, and we anticipate that the results will not vary significantly \citep{mukund2014field}. For safety purposes, the observation site is enclosed by a thin metal wire fence extending up to approximately $0.5$ m above ground level around its perimeter. The flat grassland surrounding the observation site, with grass approximately $\sim 10$ cm in height maintained by the airport authority, offers an unobstructed view for remote-sensing instruments to scan with minimal disruptions. The description of geographical details such as terrain, soil properties, vegetation, and climate around KIAB, which influence its thermodynamic and dynamic parameters (e.g., wind speed and direction, temperature and moisture in atmosphere and soil, etc.) can be found in a recent article by \citet{kutty2019fog}. 

As the observation site is located in the tropical region, convective systems  of local to large scale are commonly observed throughout the year.  In the KIAB region, it is difficult to get enough days that are completely free from clouds. Hence, for analysis, we have selected days when reported cloud cover in the Meteorological Aerodrome Report (METAR) at KIAB is $<2$ octas during the analysis period (10:00 UTC [03:30 IST] and 18:30 UTC [23:59 IST]) which we call clear-sky days. Since the field campaign at the site is an ongoing project;  we report data and analysis for a total of 80 clear-sky days from 2021-22 and 2022-23, which consists of two winter seasons (December and January) as well as two spring season (February and March). During these days, calm and clear sky conditions prevailed. In the winter season, the ABL is found to be stable with frequent occurrence of dense fog in the morning hours. Because of the prevailing easterly wind  in both the seasons, meteorological conditions at the observation site get modulated by large-scale systems developed in the Bay of Bengal (BoB), which are observed most days of the year. 

Several instruments have been deployed at the observation site to measure different parameters in the atmosphere and the soil (see Table \ref{tab: instruments}). The probe locations range from 0.5 m inside the soil to 10 km into the atmosphere (See Figure \ref{fig: observation site}c). The category of instruments, measured quantities, range, resolution, and sampling interval have been presented in Table \ref{tab: instruments}. Windcube, an active lidar-based remote sensing device, provides wind profiling in the hemispherical volume of a radius of 3 km using three different modes of scan: Doppler Beam Swinging (DBS), Range Height Indicator (RHI), and Plan Position Indicator (PPI). Wind data quality is ensured based on the carrier-to-noise ratio (CNR), which depends on the concentration and size distribution of aerosols, dust particles, clouds, and fog droplets in the atmosphere. Since emitted radiation from the windcube can not penetrate through the thick cloud/fog layer, CNR drops significantly above that layer, which leads to noisy wind data, but it detects the presence of fog/cloud above KIAB.

Temperature and humidity profiles play a key role in modulating the radiation budget in the atmosphere. Humidity and Temperature profiling (HATPRO) radiometer, a passive remote sensing device, continuously retrieves temperature and moisture profiles up to a height of 10 km using 14 channels of microwave (MW) radiation. However, moisture profiling is coarse in vertical resolution \citep{blumberg2015ground}. An Infrared (IR) sensor integrated with HATPRO retrieves the cloud/fog base, its thickness, and liquid water mixing ratio with poor accuracy. In case of heavy rain, the optical window (Radome sheet cover) of HATPRO wets and introduces noise in observed vertical profiles, but a high-temperature blower integrated with HATPRO dries the optical cover just after rain, and the data quality gets restored. LN2 calibration of HATPRO was performed periodically to ensure quality data of the temperature and humidity profiles. 

An automatic weather station (AWS) integrated with HATPRO gives temperature, pressure, relative humidity (RH), rain rate, wind speed, and direction at 2 m above ground level (AGL). Observations from AWS and vertical profiles from the HATPRO are used to retrieve many thermodynamic parameters such as water vapor mixing ratio, total precipitable water, liquid water mixing ratio, liquid water path, dew point temperature profile, different stability indices such as convective available potential energy (CAPE) and convective inhibition energy (CINE). Although HATPRO misses and misplaces the liquid water mixing ratio in the vertical direction and gives poor profiling of the liquid water mixing ratio, the windcube detects the cloud and fog base height based on the sudden change in CNR. The occurrence of cloud/fog is also well detected through a sharp change in the incoming longwave radiation from the 4-component radiation sensor (Discussed in Section \ref{Sec: Result_section}). 

Two humidity sensors and two radiation sensors, integrated with two internal Pt100 temperature sensors, are installed at 1.14 m and 1.93 m heights on a 2 m vertical mast. Note that the height of the mast is limited to 2 m due to operational constraints at the airport. A 1.5 W heater, integrated with each radiation sensor, is kept on to avoid condensation on the optical window of sensors. However, radiation sensors are not integrated with ventilated units, which can introduce errors up to $\pm$ 15 W m$^{-2}$ whenever natural ventilation is not sufficient. Additionally, twenty temperature sensors are mounted on the same mast, enabling high vertical resolution measurements from 4.5 cm to 2 m above ground level (AGL).  These sensors are calibrated in an isothermal bath, and the maximum relative errors among the sensors are $<0.1$ K. The high vertical resolution close to the surface is intended to capture the Lifted Temperature Minimum (LTM) height and its intensity. To obtain a temperature profile in the soil and measure sensible heat flux (SHF) at the soil surface, a soil temperature profiler and two heat flux sensors are installed near the concrete base. All sensors, including those on the mast and within the soil, are connected to a data logger (Keysight DAQ970A) to record the measurements continuously. Windcube, HATPRO, and the data logger are connected through three mini-computers with uninterrupted internet connections, facilitating continuous remote monitoring. To ensure a continuous and stable power supply, all computers and other accessories are installed in two IP65 electrical enclosures (each measuring 0.5 m × 1.0 m × 1.0 m) located near the concrete base and connected to the reliable airport's electricity supply. 

%%%%%%%%%%%%%%%%%%%%%%%%%%%%%%% Table %%%%%%%%%%%%%%%%%%%%%
{\small
\begin{longtable}{p{2cm}p{3cm}p{3cm}p{4cm}}
\multicolumn{4}{c}{Details of Instruments Deployed}\\
\hline
    \rowcolor{Gray}
     Categories & Instruments &  Measured quantities & Range (RA), Sampling interval (SA) and resolution (RE)\\
\hline
    
     \multirow{2}{*}{Soil sensors} & Soil temperature profiler (STP01, Hukseflux) &  Soil Temperature at 0.02, 0.05, 0.1, 0.2 and 0.5 m depth of soil
     %with $\pm$  0.005m accuracy from soil top
     & RA: -30 to 70$^{\circ}$C, Absolute uncertainty: $\pm$0.7 K, relative uncertainty: $\pm$0.05 K, SA: 5 sec\\
% \hline
    & Soil Heat flux (HFP01, Hukseflux) &  Heat flux at 0.05 m depth of soil %with $\pm$  0.005m accuracy from soil top
    & RA: -2000 to 2000 W, uncertainty: $\pm$3 $\%$, SA: 5 sec\\
\hline
    \rowcolor{Gray}
    \multirow{2}{*}{\parbox{2cm}{\vspace*{1mm} Surface layer within 2 meters AGL}} & 20 thermistors (Te connectivity sensor NTC Discrete MBD 10 kilo-ohm) &  Temperature & 
     RA: -40 to 125$^{\circ}$C, uncertainty: $\pm$0.2$^{\circ}$ between 0--70$^{\circ}$C, SA: 5 sec\\
% \hline
    \rowcolor{Gray}
    & 2 humidity sensors (HIH-5030/5031 series, Honeywell) &  RH & RA: 0 to 100$\%$, uncertainty: $\pm3\%$ from 11--89$\%$, otherwise $\pm7\%$, SA: 5 sec\\
\hline
%{\color{red}
    \multirow{6}{*}{\parbox{2cm}{Weather sensors \\ at 2 m AGL}} & \multirow{6}{*}{\parbox{2cm}{Multi-component weather sensors (Vaisala, WXT530 Series)}} & Air temperature & 
     RA: -52$^{\circ}$C--60$^{\circ}$C, uncertainty: $\pm$0.3$^{\circ}$C, SA: 1 sec\\
     
    &  & RH & RA: 0--100 $\%$, uncertainty: $\pm 3\%$ at 0--90$\%$ \& $\pm5\%>90\%$, RE: 0.1$\%$, SA: 1 sec\\
    
    &  &  Barometric pressure & RA: 600 to 1100 hPa, uncertainty: $\pm$0.5 hPa, RE: 0.1 hPa, SA: 1 sec\\
    
    &  & Precipitation & 
     RA: 0 to 200 mm/h, uncertainty: $\pm5\%$, SA: 10 sec\\

    &  & Wind speed & RA: 0 to 60 m/s, Accuracy: $\pm3\%$ at 10 m/s, RE: 0.1 m/s, SA: 1 sec\\
    
    &  & Wind direction & 
     RA: 0 to 360$^{\circ}$, accuracy: $\pm$3$^{\circ}$ at 10 m/s, RE: $1^{\circ}$, SA: 1 sec\\
 %   }
\hline
    \rowcolor{Gray}
    Radiative fluxes & 4-component net radiometer (NR01, Hukseflux) & Radiative flux (upward LW and SW, downward LW and SW) & Calibration uncertainty solar: $<1.8\%$, calibration uncertainty longwave: $<7\%$, SA: 5 sec\\
\hline
    \multirow{6}{*}{\parbox{2cm}{Atmospheric profiles}} & Wind lidar (Windcube 100S, Leosphere) & Wind speed and direction in the hemisphere of 3 km radius & RA: -30--30 m/s in the radial direction, accuracy: $\pm0.5\%$, RE: 0.1$^{\circ}$ resolution, range resolution: 50 m, SA: 20 sec to 3 minutes (based on mode of scan)\\

    & Humidity and temperature profiler (HATPRO, Radiometer Physics, A Rohde \& Schwarz Company) & Profile of temperature, RH, water and liquid water mixing ratio, ABL height, cloud base height, and stability profiles & Total 93 vertical levels from 10 m AGL to 10 km, having a resolution of 25 m to 300 m, RMS accuracy of water vapor mixing ratio: $\pm0.3$ g m$^{-3}$, Temperature accuracy: $\pm$0.25 K with 500 m, boundary layer ($<2$ km) mixing ratio accuracy: $\pm$ 0.03 g m$^{-3}$, SA: 60 sec\\
\hline
\caption{Details of instruments deployed at the observation site (KIAB) and description of different meteorological variables} 
\label{tab: instruments} 
\end{longtable}  }
To study the effect of fog on LTM and radiation divergence, fog data is taken from the METAR, an airport weather monitoring report used for aviation purposes \cite{mesonet2019asos}. Since the METAR station is located $\sim$1 km east from our observation site and METAR reporting is half-hourly, there exists a chance of temporal offset in the reporting of fog by METAR and the corresponding response of sensors at our observation site. In this paper, 5-minute averaging has been performed on all data except METAR to avoid spurious observations. Unless stated otherwise, all heights are reported relative to the local ground level.

%%%%%%%%%%%%%%%% Model %%%%%%%%%%%%%%
\section{Radiation model with aerosols} \label{sec: rad_model}
The radiation model used in this study is the modified version of the band model used by \citet{varghese2003fast}, which itself is adopted from the band model developed and improved by \citet{chou1993one, chou1994efficient, chou2001thermal}. Later, \citet{edwards2009radiative, ponnulakshmi2012hypercooling, ponnulakshmi2013hypercooling} pointed out the erroneous assumption of the Planckian nature of downward longwave (LW) radiation in the reflected radiation term used in the model by \citet{varghese2003fast} (which results in a spurious source of cooling near the surface, having a nonphysical length scale). We take the model with the corrections suggested by \citet{ponnulakshmi2012hypercooling, ponnulakshmi2013hypercooling}. In the modified version of the model, downward and upward radiative flux divergence at height $z$ is given by
\begin{equation}
    \frac{dF^{\downarrow}_{ji}}{dz} = -A^i_j[c_i^j\pi B_j(T) - F_{ji}^{\downarrow}]
    \label{eq: downward_flux_divergence}
\end{equation}

\begin{equation}
    \frac{dF^{\uparrow}_{ji}}{dz} = A^i_j[c_i^j\pi B_j(T) - F_{ji}^{\uparrow}]
    \label{eq: upward_flux_divergence}
\end{equation}

The top boundary condition comes from the fact that there is no incoming longwave radiative flux at the top of the atmosphere, and fluxes for the bottom boundary are given by reflected and emitted components of longwave radiation from the ground.
\begin{equation}
    F_{ji}^{\downarrow}(\infty) = 0
    \label{eq: top_boundary_condition}
\end{equation}

\begin{equation}
    F{ji}^{\uparrow}(0) = c_i^{j}[\epsilon_s \pi B_j(T_s)] + (1-\epsilon_s)F_{ji}^{\downarrow}(0)
    \label{eq: bottom_boundary_condition}
\end{equation}
where $F_{ji}$ is the radiative fluxes for sub-band $i$ of band $j$. $A_j^i$ is given by
\begin{equation}
    A_j^i = dk_j^i \bigg(\frac{P}{P_r}\bigg)^m f_j(T, Tr)\rho_w
\end{equation}
The value of $c_i^j$, $d$, $k_j^i$, $P_r$, m, and $T_r$ as well as pressure and temperature scaling have been taken from \citet{chou1994efficient, chou2001thermal}. $B$ is the Planck function of radiation. $T_s$ is the surface temperature of the surface having emissivity $\epsilon_s$.

\subsection{Inclusion of aerosol-radiation interactions} \label{sec:aerosol_model}
The vertical profile of the aerosols close to the ground plays a key role via a change in radiative flux divergence \cite{mukund2010hyper, mukund2014field}. From laboratory experiments, \citet{mukund2014field} showed that the LTM intensity decreases with decrease  in aerosol concentrations in the test section, and it disappears when the aerosol concentrations in the test section are reduced significantly by filtering aerosols or by blocking radiation interaction in the test section with a thick opaque sheet. Noting the experimental observations of \citet{mukund2014field}, we have developed an extension to the corrected version of the model \citet{ponnulakshmi2012hypercooling, ponnulakshmi2013hypercooling} by including aerosol-radiation effects.  The extent to which a spherical aerosol particle ($s$) of radius $r$ and refractive index $n+k\textit{i}$, interacts with radiation at wavenumber $\nu$ is given by the extinction cross-section $\sigma_{ext}^{s}(\nu, r)$ \cite{stephens1984parameterization, jacobson1999fundamentals}.  Effect of the hygroscopic growth of aerosols on radiation has been accounted via a change in refractive index when RH changes \citep{hess1998optical}. For simplicity, the shape of aerosols is considered to be spherical, and independent scattering dominates for typical aerosol concentrations in the atmosphere \cite{liou2002introduction}. Under these conditions, extinction efficiency ($\sigma_{ext}^{s}(\nu, z)$) at wavenumber $\nu$ for an aerosol species $s$ having distribution $N_s(r, z)$ at height $z$ is given by
\begin{equation}
    \sigma_{ext}^{s}(\nu, z) = \sum_{r} \sigma_{ext}^{s}(\nu, r) N_s(r, z)
\end{equation}
where $N_s(r, z)$ follow log-normal distribution at height 
 $z$ \cite{hess1998optical} and is given by 
\begin{equation}
    \frac{dN_s(r, z)}{dr} = \frac{N_s(z)}{\sqrt{2\pi} r \log\sigma_i \ln{10}}exp\bigg[-\frac{1}{2}\bigg(\frac{\log r - \log r_{mod N,s}}{\log \sigma_i}\bigg)^2\bigg]
    \label{eq: log-normal_distribution}
\end{equation}

where $N_s(z)$ is the total number density of aerosol species $s$ at height $z$; $\sigma$ and $r_{modN, s}$ are distribution parameters. $\sigma_{ext}^{s}(\nu, r)$ is calculated from the standard BHMIE code \cite{bohren2008absorption} and $\sigma_{ext}^{s}(\nu, z)$ is summed for all aerosol species ($s$) and sub-band interval $i$ to get total extinction efficiency in band $j$ at height $z$
\begin{equation}
    \sigma_{ext}^{j}(z) = \sum_{\nu=\nu_{ij}}\sum_{s=1}^{S}\sigma_{ext}^{s}(\nu, z)
\end{equation}

The diffuse transmission function for the aerosol in band $j$, at level $z$ is given by \citet{liou2002introduction}
 \begin{equation}
     \tau_j^{aer}(z) = exp\bigg(-d\int_0^z \sigma_{ext}^{j}(z')dz'\bigg)
    \label{eq: tau_aerosol}
\end{equation}
When water vapor and aerosols are both present, the transmission function for a band $j$ is given by:
\begin{equation}
    \tau_j(eff) = \tau_j^{wv}\tau_j^{aer}
    \label{eq: tau_eff}
\end{equation}
where $\tau_j^{wv}$ is given by  
\begin{equation}
    \tau_j^{wv} = \sum_{i=1}^{m_j} c_i^j\tau_j^i
    \label{eq: tau_water_vapor}
\end{equation}
where $\tau_j^i$ is given by 
\begin{equation}
    \tau_j^i = exp\bigg[-d\int_0^z k_j^i \bigg(\frac{P}{P_r}\bigg)^m f_j(T, T_r)\rho_w dz'\bigg]
    \label{tau_wv_ji}
\end{equation}

From Eqs. (\ref{eq: tau_aerosol}, \ref{eq: tau_eff}, \ref{eq: tau_water_vapor}, \ref{tau_wv_ji}), $\tau_{j}^i(eff)$ is given by 

\begin{equation}
    \tau_{j}^i(eff) = exp\Bigg[-d \int_0^z \bigg(\sigma_{ext}^{j}(z')+k_j^i \bigg(\frac{P}{P_r}\bigg)^m f_j(T, T_r)\rho_w\bigg)dz'\Bigg]
    \label{eq: tau_j_i_eff}
\end{equation}

To account for the role of aerosols, $A_j^i$ is updated in the model by using Equation (\ref{eq: tau_j_i_eff}) as 
\begin{equation}
    A_j^i = d\bigg[\sigma_{ext}^{j}(z)+k_j^i \bigg(\frac{P}{P_r}\bigg)^m f_j(T, T_r)\rho_w\bigg]
    \label{eq: A_j_i_redefined}
\end{equation}
Inserting the updated expression of $A_j^i$ from Equation (\ref{eq: A_j_i_redefined}), we get the updated radiation model, which accounts for aerosol-radiation interactions. In this updated model, Equation (\ref{eq: downward_flux_divergence}, \ref{eq: upward_flux_divergence},  \ref{eq: bottom_boundary_condition} and \ref{eq: tau_water_vapor}) remains the same along with the updated Equation (\ref{eq: A_j_i_redefined}).

From Equation (\ref{eq: downward_flux_divergence}) and (\ref{eq: upward_flux_divergence}), total radiative flux divergence at level z is given by 
\begin{equation}
    \frac{dF}{dz} = \sum_i \sum_j \bigg(\frac{dF_{ij}^{\uparrow}}{dz} - \frac{dF_{ij}^{\downarrow}}{dz}\bigg)
    \label{eq: total_radiative_flux_divergence}
\end{equation}
Equation (\ref{eq: total_radiative_flux_divergence}) becomes the source term in the 1-dimensional radiation-conduction equation which is given by:
\begin{equation}
    \frac{\partial T(t, z)}{\partial t} = \alpha \frac{\partial^2 T(T, z)}{\partial z^2} - \frac{1}{\rho C_p}\frac{dF}{dz}
    \label{eq: 1_dim_cond_radiation_model}
\end{equation}
where $z\in (0, H)$ and $t>0$. $H$  is the height of the atmosphere, and $\alpha$ is the molecular thermal diffusivity of air. Equation (\ref{eq: 1_dim_cond_radiation_model}) is solved using the Thomas Algorithm \cite{press2007numerical} to get the temperature and radiative flux evolution in the atmosphere in the presence of water vapor and aerosols.

%%%%%%%%%%%%%%%%%%%%%%%
\subsection{Aerosol concentrations and profile} \label{sec: aerosol_con_and_prof}
Due to the unavailability of aerosol data at the observation site, representative aerosol properties have been taken from the OPAC (Optical Properties of Aerosols and Clouds) database to account for the role of aerosols in the model \cite{hess1998optical}. Although this database contains diverse aerosol profiles for different atmospheric conditions, the considered radiation model has been integrated with urban aerosols (which consists of insoluble; INSO,  water-soluble; WASO and soot; SOOT particles) because flight operations and other construction-work close to the observation site emit a massive amount of aerosol particles in the atmosphere.  Note that only WASO particles show hygroscopic growth as a function of RH \citep{hess1998optical}, which has been accounted for in all simulations. The parameters for the size distribution and refractive index for the different aerosol components   at different RH have been taken from the same database where all components follow log-normal distribution as shown in Equation (\ref{eq: log-normal_distribution}). 

\citet{hess1998optical} has considered the roughly uniform concentration of aerosols (scale height of 8 km) in the ABL, which has not been observed in Indian tropical regions \cite{devara1993lidar, devara1995real, chate2004field}. \citet{devara1993lidar} and \citet{devara1995real} have reported a decrease in the aerosol concentration with an increase in height (measured above 40 m AGL) from long-term lidar observations at a tropical site. \citet{chate2004field} has shown uniform aerosol concentration within 1 m AGL in the range of  $10^{4}$--$10^{5}$ cm$^{-3}$, which is 10-100 times higher compared to the concentration measured above ~40 m. Hence, we use the aerosol concentration profile, based on the measurements by \citet{devara1993lidar, devara1995real, chate2004field}, and we use the properties of aerosol species as provided by the OPAC database \cite{hess1998optical}.

\citet{mukund2010hyper} fitted the Rouse profile to the aerosols profile from \citet{devara1993lidar} and arrived at the functional profile as in Equation (\ref{eq: rouse_profile}) for the variation of aerosol number density with height. Taking into account the observations of \citet{chate2004field} and \citet{mukund2014field} calculations, we use different equations for  aerosol concentrations above 1 m and below 1 m differently (Equation \ref{eq: rouse_profile}, \ref{eq: 1m_below1} and \ref{eq: 1m_below2}) 
\begin{equation}
    N_i(z) = N_{i0} \bigg(\frac{z}{Z}\bigg)^{-p}, z>1m
    \label{eq: rouse_profile}
\end{equation}

\begin{equation}
    N_i(z) = N_c exp(-z+1), z\le1m
    \label{eq: 1m_below1}
\end{equation}

\begin{equation}
    N_c = N_{1m+} (1.0+10^{-6})
    \label{eq: 1m_below2}
\end{equation}

where $p=0.74$ and $z = 50$ m and $N_{i0}$ is total concentration at height $z$ \cite{mukund2010hyper, hess1998optical}.

%%%%%%%%%%%%%%%%%%%%%%%%%%%% graph %%%%%%%%%%%%%%%%%%%%%%%%%%%%
\begin{figure}[ht]
	\centerline{\includegraphics[width=0.5\textwidth]{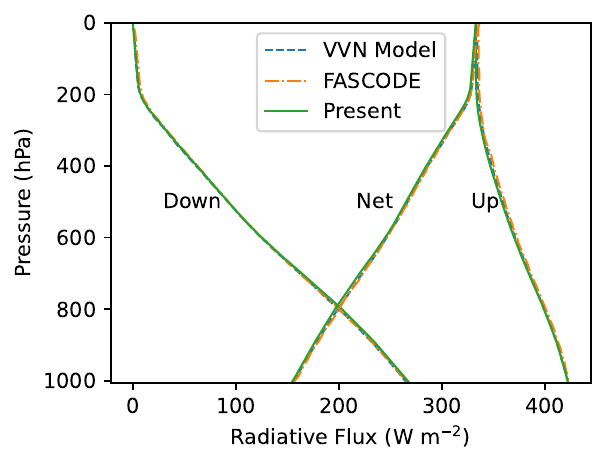}}
        \caption{Comparison of upward, downward, and net flux, with the radiation model by \citet{varghese2003fast}, (this model produces correct result at $\epsilon_s = 1$) and FASCODE \cite{clough1992line} for $\epsilon_s = 1$. In these models, Midlatitude summer atmosphere with water vapor absorption has been used to calculate fluxes.}
	\label{fig: validation}
\end{figure}

\begin{figure}[ht]
	\centerline{\includegraphics[width=0.6\textwidth]{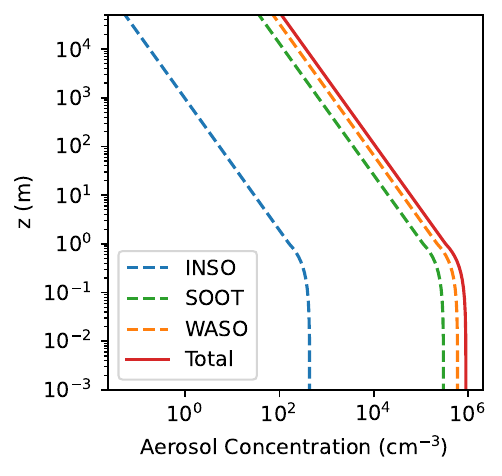}}
	\caption{ Vertical profile of aerosol concentration (accumulation and coarse mode only): concentration of water-soluble (WASO) and soot particles (SOOT) are significantly higher than that of insoluble particles (INSO), and concentrations are roughly uniform below 1 m AGL.} 	
	\label{fig: aerosol concentration} \vspace*{-2mm}
\end{figure}

Since direct measurements of aerosol concentration profile and its properties are not available at the observation site, we have tested 1 to 7 times of the reported concentration in \citet{hess1998optical} in the present work to account for the uncertainties in spatial and temporal variability of aerosol particle concentrations and properties \cite{hess1998optical, wang2011multi, forster2021earth, wu2021urban}. If we take the same aerosol concentrations as reported in \citet{hess1998optical}, the steady-state simulated LTM intensity is less than 0.2 K, which is weaker compared to our field observations as well as other field observations \citep{mukund2014field}. Instead, if we consider six times (6x) higher concentration of aerosols compared to the reported concentration in \citet{hess1998optical} (Figure \ref{fig: aerosol concentration}), the LTM and LHR obtained from the simulations are in good agreement with our field observations.  Note that aerosol particles of diameter less than 0.1 $\mu$m (ultra-fine particles) are not relevant to optical interactions \citep{seinfeld2016improving}, hence aerosol concentration presented here consists of accumulation and coarse mode only. With this profile, the total aerosol concentration at 1 m AGL is 3.3$\times 10^{5}$ cm$^{-3}$, whereas it is 2.15$\times 10^{4}$ cm$^{-3}$ at 40 m AGL, which is in the same order of magnitude, reported from many field observations \cite{wu2021urban}. Hence, we will use aerosol concentration profiles shown in Figure \ref{fig: aerosol concentration} for all further analysis. In Section \ref{sec: discussion}, we will discuss the implications of higher concentrations of aerosols close to the ground.

\subsection{Validation of the model}
Although the radiation model by \citet{varghese2003fast} produces a spurious cooling near the ground due to incorrect handling of reflected term, such spurious cooling does not occur for $\epsilon_s = 1$ because the reflected component of downward radiative flux vanishes (Equation \ref{eq: bottom_boundary_condition}). Hence, the present model has been validated against the model by \citet{varghese2003fast} and the line-by-line Fast Atmospheric Signature Code (FASCODE) by  \citet{clough1992line} for $\epsilon_s = 1$ for the Mid Latitude Summer (MLS) standard atmosphere with water vapor line absorption only. Figure \ref{fig: validation} shows the comparison of upward, downward, and net flux with the radiation model by \citet{varghese2003fast} model and FASCODE. At the top of the atmosphere, offset in upward fluxes in the present model goes up to 7 W m$^{-2}$ as compared to FASCODE, and a similar offset in upward fluxes has also been observed between the model by \citet{varghese2003fast} and FASCODE \cite{varghese2003fast}. Near ground, the relative offset of fluxes in the present model is less than 2 W m$^{-2}$ compared to the model by \citet{varghese2003fast} and FASCODE. 

\subsection{Initialization of radiation model} \label{subsec: init_radiation_model} 
Vertical profile of temperature and water vapor mixing ratio with respect to pressure and height and aerosols profiles are required to initialize the model. To get the temperature and humidity profile from the surface to 50 km AGL, data from three different sources have been concatenated according to the height: surface to 2 m data from the mast observations, 2 m to 10 m data from MW radiometer measurements, and 10 km to 50 km from spatially interpolated ERA5 reanalysis dataset \cite{era5_data}.   Since, true vertical resolution of the retrieved water vapor profile is coarse \citep{blumberg2015ground}, we have performed sensitivity analysis by $\pm$20\% change in mixing ratio at all heights ($\pm$1.2 g/kg change at 10 m AGL), magnitude of change in temperature is less than 0.2 K  at any height. It is one order of magnitude smaller compared to observed LTM intensity (Section \ref{fig: ltm_vertical_profiles}),  also variation of the mixing ratio alone in this range will not produce an LTM type profile. Hence, we will use the observed mixing ratio profiles from the MW radiometer. To resolve LTM and radiation divergence, the vertical resolution of the present model is kept at 0.4 mm within 1 m AGL, and later, it gradually coarsens to 18 m at 50 km AGL, which counts a total of 32771 vertical grid points in the model. All simulations have been performed with and without aerosol profiles for all the considered days at this resolution.

 For the bottom boundary condition, we have used observed/estimated surface temperature ($T_s$). We don't have a direct measurement of $T_s$ for the initial 44-days of analysis, and hence, we have estimated it using the radiosity Equation (\ref{eq: surf_temp_modelling}), whenever required. The estimated surface temperature has been validated against the observed surface temperature (mounted later in January 2023), which is available for the last 36-days of the analysis. Here, it is to be noted that with a surface emissivity of $\epsilon_s = 0.95$, the mean and standard deviation of the absolute error between estimated and observed $T_s$ from local sunset to sunrise time is 0.42 K and 0.3 K, respectively. This discrepancy is 5--6 times smaller than the observed LTM intensity derived from direct $T_s$ observations.(See supplementary figure S1).   
\begin{equation}
    F^{\uparrow}_s = \epsilon_s \sigma T_s^4 + (1-\epsilon_s)F^{\downarrow}_s
    \label{eq: surf_temp_modelling}
\end{equation}
where $F^{\uparrow}_s$ and $F^{\downarrow}_s$ are upward and downward LW flux respectively. 

Each day, simulation begins with a concatenated temperature and water vapor mixing ratio profile, initiated two hours before local sunset as a model spin-up. After 5--6 hours post local sunset, mist forms, reducing visibility to less than 5 km for most days (Figure \ref{fig: LTM observations}a). This can lead to sensor wetting and measurement errors when relative humidity (RH) exceeds 80\%. Therefore, analysis and simulation are primarily restricted to within 4 hours after local sunset when RH remains below 75\%. However, when discussing fog/cloud effects on LTM (Subsection \ref{sec: fog_thickness}), mast-temperature observations during fog are approached with caution. Significant changes in CNR at 50 m AGL occur with mist or fog presence, attributed to variations in micron-sized particle (water droplet) concentration or size distribution not considered in the current model. As aerosol concentrations remain constant over time in simulations, analysis is limited to 4 hours post local sunset to mitigate observed mist or fog effects at the site and condensation-induced measurement errors.

\section{Results}\label{Sec: Result_section}
\subsection{LTM observations}
A temperature profile within 2 m AGL is considered an LTM profile if the minimum temperature between 10 cm to 55 cm AGL is lower than the surface temperature and the temperature between 1.6 m and 2 m  by at least 0.3 K. Note that in the absence of direct observations for ground surface temperature, we have used temperature measured at 4.5 cm to quantify LTM occurrence. Threshold of 0.3 K is chosen to avoid the spurious observation of LTM  
whereas the threshold for upper and lower limit of LTM height is taken from \citet{blay2015lifted} and \citet{oke1970temperature}.

\subsubsection{LTM occurrence and characteristics}
\begin{figure}[ht]
	{\centerline{\includegraphics[width=0.9\textwidth]{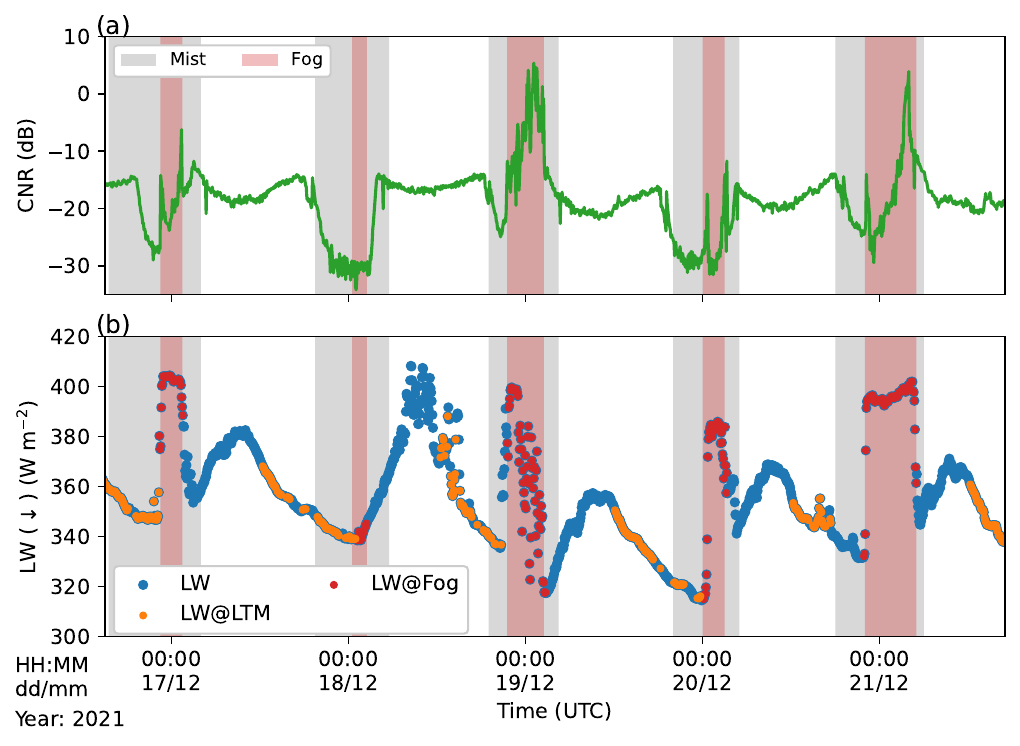}}}
	\caption{Observations of LTM over a few days (a) Change in CNR at 50 m AGL in the presence of mist and fog indicates a change in concentration or distribution of micron-size water droplets in the atmosphere. (b) LTM appears during evening transitions and maintains for hours before fog occurs. LTM is not observed during day hours, local conditions are based on METAR notification and reported minimum-visibility on 17$^{th}$, 18$^{th}$, 19$^{th}$, 20$^{th}$ and 21$^{st}$ December , 2021 are 96 m, 800 m, 193 m, 48 m and 48 m respectively.}
	\label{fig: LTM observations}
\end{figure}

 Although it has been speculated that LTM would not appear in cloudy conditions or it disappears when cloud/fog appear \cite{mukund2014field, blay2015lifted}, the behavior of LTM due to change in the downward LW flux in foggy conditions has not been reported quantitatively through observations. From the simultaneous measurement of LW flux during LTM occurrences and fog events from our field experiments, we show that downward LW flux changes substantially with the appearance of fog (Figure \ref{fig: LTM observations}b). In this plot, downward LW flux obtained from the radiation sensor mounted at 1.14 m AGL has been overlapped with LTM (orange marker) and fog occurrence (red marker) to observe LTM behavior in different atmospheric conditions. Since on all considered days show similar behavior (for eighty days, covering two fog seasons), data from only five days have been shown for brevity. It can be observed that downward LW flux shows a distinctive diurnal variation (except when fog appears), indicating that the sky is free from convective system. Also, the precipitation sensor has not recorded any precipitation (not shown). However, mild fluctuations in the downward LW flux during the local afternoon on some days indicate the presence of fair-weather cumulus clouds associated with local convection. 

We note that, LTM is absent during the daytime but emerges as the evening transition progresses, persisting for several hours on clear nights or intermittently if local conditions are unfavorable.  When fog develops in the early morning, LTM disappears, which is also accompanied by a sharp increase in downward longwave (LW) flux (except on 18$^{th}$ December 2021). It should be noted that temperature from the mast sensors and, hence, LTM occurrence might not be accurate because of sensor wetting, but a similar observation, like disappearance of LTM in the presence of fog/cloud has been reported in other field experiments \citep{mukund2014field, blay2015lifted}. We will further look at this aspect from idealized simulations in Section \ref{sec: fog_thickness}.  From METAR observations, fog reported on $18^{th}$ December 2021 was mild, having a minimum visibility of 804 m (possibly an optically thin layer of fog). Because of the thin layer of the fog, downward LW flux as well as divergence does not change sharply, and LTM persists during this fog event. When sunlight causes the dissipation of fog in the morning, downward LW flux, as well as radiation divergence, returns to its diurnal cycle. However, because of solar heating, local convection close to the surface dominates and does not allow sustenance of LTM. Hence, LTM is not observed during daytime in spite of the sky being cloud-free.

Based on eighty days of observations spanning two winter and two spring seasons, we find that LTM characteristics remain consistent across seasons. During winter, LTM intensity averages $2.3\pm0.7$ K, while in spring it is $2.0\pm0.5$ K (mean and one standard deviation). Likewise, LTM height measures $0.30\pm0.10$ m in winter and $0.32\pm0.12$ m in spring. Overall, across the eighty-days analysis period, LTM intensity averages 2.2$\pm$0.6 K, with a height of 0.31$\pm$0.11 m. These findings align well with previous studies (\citet{mukund2014field, blay2015lifted}). Our observations indicate slightly higher LTM intensity compared to \citet{blay2015lifted} and lower than \citet{mukund2014field}. Minor differences in LTM characteristics across different field experiments may stem from variations in favorable conditions such as calm, clear skies, ground properties, and aerosol characteristics.

\subsubsection{Parameters that control LTM}\label{subsection: parameters_ltm}

\begin{figure}[ht]
	\centerline{\includegraphics[width=0.8\textwidth]{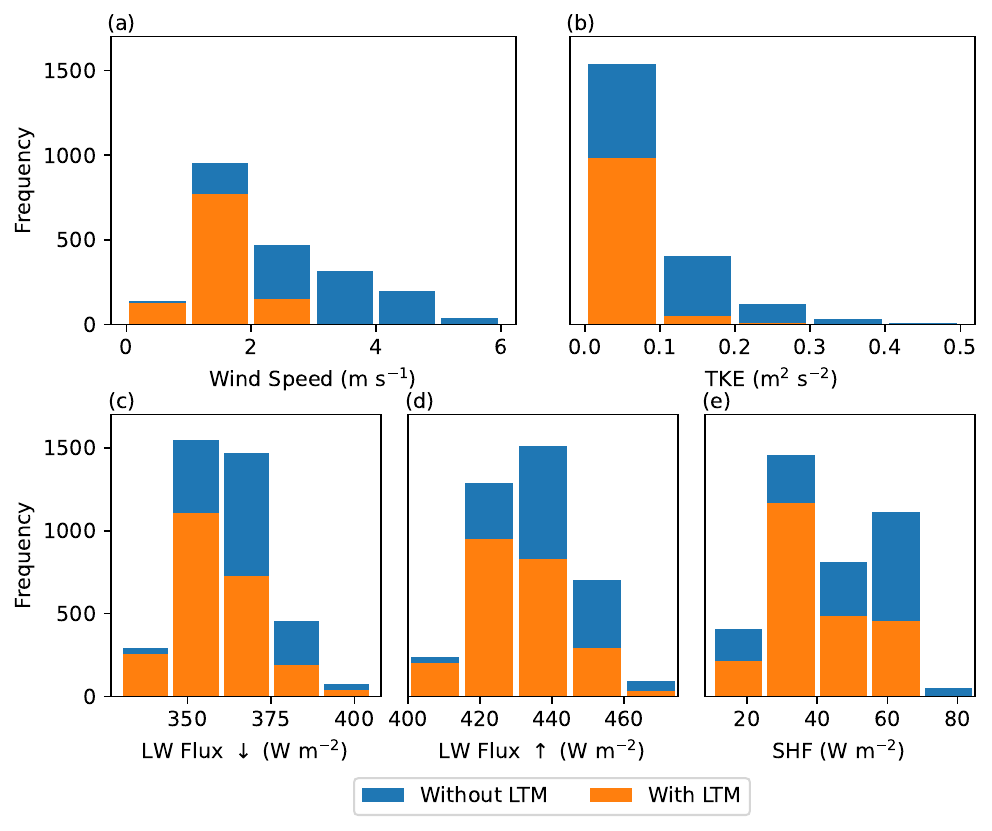}}
	\caption{  Histogram of different meteorological factors with and without LTM. (a) Wind speed and (b) TKE at 2 m AGL;
 (c) downward and (d) upward LW flux at 1.14 m AGL; (e) Surface sensible heat flux (SHF) at 0.05 m below ground.}
	\label{fig:flux_tke_and_ltm}
\end{figure}

 Calm and clear sky conditions are favorable for LTM occurrence, which implies that wind speed, TKE, and downward and upward LW flux are important parameters that control the LTM characteristics \cite{lake1956temperature, mukund2014field, blay2015lifted}. Further, Apart from aerosol characteristics near the ground, SHF at the soil surface controls the surface temperature evolution and hence, LTM intensity. Histogram of these meteorological factors with and without LTM occurrence from local sunset to the next 4 hours are shown in Figure \ref{fig:flux_tke_and_ltm}. Unfortunately, since the wind sensor was not functioning during the winter and spring seasons of 2022-23, we have used initial 44 days of wind data from season 2021-22 for analysis (Figure \ref{fig:flux_tke_and_ltm}a and \ref{fig:flux_tke_and_ltm}b).   

Low wind speed and fluctuations introduce minimal disturbances so that an LTM profile can sustain. Wind speed measured 2 m above the ground level is  $<$ 6 m s$^{-1}$ indicate the calm condition during the observation period (Figure \ref{fig:flux_tke_and_ltm}a). We observe that LTM occurrence strongly depends on the wind speed. Most ($>$ 80 \%)  of the LTM occurrence is when the wind speed is less than 2 m s$^{-1}$. Moreover, LTM does not appear or sustain if wind speed is more than 3 m s$^{-1}$. A drop in frequency of LTM occurrence with the increase in wind speed rules out the role of advection or drainage flow in LTM development, which is also reported in other field experiments \cite{mukund2014field, blay2015lifted}.

Apart from mean wind characteristics, turbulence is another key parameter in LTM development \cite{narasimha1994dynamics, mukund2014field, blay2015lifted, jensen2016observations}. Since turbulent kinetic energy (TKE) is a quantitative measure of the intensity of turbulence, we calculate it from horizontal wind measured at 2 m AGL using Equation \ref{eq: tke}.
\begin{equation}
    TKE = \frac{1}{2} \big(\overline{u'^2} + \overline{v'^2}\big)
    \label{eq: tke}
\end{equation}
where $u$ and $v$ are easterly and northerly components of wind, respectively, sampled at the interval of 1 second. Velocity fluctuation is calculated as $u' = u-\overline{u}$, $v' = v-\overline{v}$, and all averaging have been done over 5 minutes. Figure \ref{fig:flux_tke_and_ltm}b shows TKE variation with and without LTM. Although TKE varies up to 0.5 m$^2$ s$^{-2}$, TKE $>$0.2 m$^2$ s$^{-2}$ is observed for less than 8\% of the observations, which indicates that the boundary layer near the ground is not highly turbulent. Further, note that most ($\sim$ 95 \%) of the LTM occurrence  is when the TKE is $<0.1$ m$^2$ s$^{-2}$ and LTM is not observed when TKE is $>$ 0.3 m$^2$ s$^{-2}$. It signifies the requirement of low-turbulent conditions within a few meters from ground for the occurrence of LTM. Since low wind speed and TKE provide favorable conditions for LTM occurrence but do not interact directly with LTM evolution, we observe that LTM intensity is poorly correlated with wind speed and LTM intensity. 

 Figures \ref{fig:flux_tke_and_ltm}c and \ref{fig:flux_tke_and_ltm}d show  histograms of observed radiative fluxes with and without LTM. Unlike wind-speed and TKE, we note that LTM appears for all observed values of LW radiative fluxes. However, the frequency of LTM occurrence increases with the decrease in upward and downward LW fluxes. It indicates that LTM occurrence, not only depends on cloud-free sky, but it also depends on the state of the local atmosphere, which can modulate the incoming LW fluxes. These factors include vertical distribution of water vapor and thermal state of the atmosphere, which can change the incoming LW radiation depending on how it is distributed vertically. Since LTM intensity depends on radiation divergence, not on radiative fluxes (Subsection \ref{sec: ltm_model_and_obs}), we observe that LTM intensity and LW fluxes are weakly anti-correlated ($r < -0.2$).

  Sensible heat flux (SHF) and surface temperature exhibit a strong correlation and are pivotal in determining the occurrence and characteristics of the LTM. Figure \ref{fig:flux_tke_and_ltm}e illustrates that the occurrence of LTM decreases with an increase in SHF. When LTM is observed, assuming other fluxes remain constant, a reduction in SHF diminishes the surface cooling rate, potentially resulting in a relatively higher surface temperature, thus intensifying LTM. We find a moderate negative correlation between SHF and LTM intensity ($r = -0.36$, p-values $<$ {\it{.001}}). Similar relationships between surface SHF and radiative cooling have been documented by \citet{gentine2018coupling}. In summary, LTM occurrence tends to increase with decreasing wind speed, Turbulent Kinetic Energy, upward and downward Longwave fluxes, as well as sensible heat flux.

\subsection{LTM in 1-D model and comparison with field observations}\label{sec: ltm_model_and_obs}

\begin{figure}[ht]
	\centerline{\includegraphics[width=0.8\textwidth]{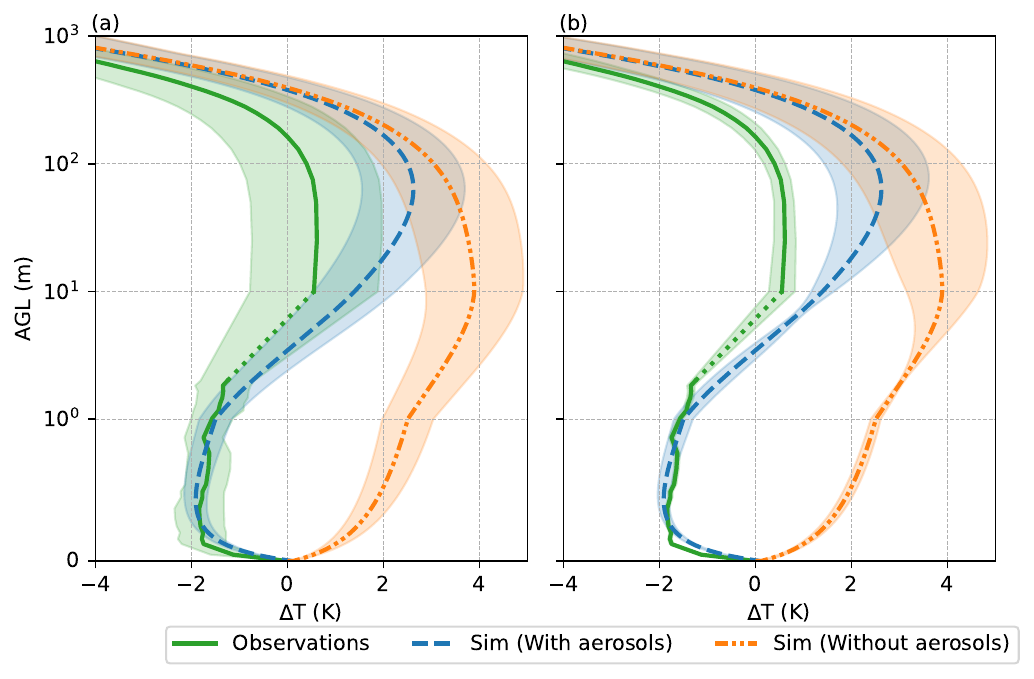}}
	\caption{ Mean relative temperature profiles (line plots) from observations (green) and simulations with aerosols (blue) and without aerosols (orange). Shading represents ($\pm 1 \sigma$) variability in the profiles for each cases. (a)  day-to-day variability in 4-hourly mean profiles (from local sunset to next 4-hours) observed over 80 days spanning two fog-seasons and (b) temporal variability of the vertical temperature profile observed during the 4 hours period after the local sunset for the same dataset. The green, dotted lines in observations represent the transition from mast-data (up to 2 m) to microwave radiometer data beyond 10 m. Note here vertical axis has hybrid-scale, a logarithmic scale above 1 m AGL, and a linear scale from the surface to 1 m AGL.}
	\label{fig: ltm_vertical_profiles}
\end{figure}
 
\citet{mukund2014field} demonstrated through laboratory experiments the necessity of aerosols, for getting observed temperature profiles and radiative cooling. However, both laboratory and field experiments conducted by \citet{mukund2014field}, were limited to a height within two meters close to the ground. For various atmospheric phenomena, such as fog occurrence, temperature and humidity profiles extending several hundred meters above ground level are crucial. In this context, utilizing our dataset, we investigate the significance of aerosols in determining vertical temperature profiles within the nocturnal atmospheric boundary layer. For all eighty-days of observations, numerical simulations were conducted with and without aerosols, for a four-hours period from sunset. Initial and boundary conditions are derived from observations as discussed in Subsection \ref{subsec: init_radiation_model}. Given that ground-level temperatures vary diurnally and daily, we present relative temperature profiles with respect to the ground surface. Results from this analysis are depicted in Figure \ref{fig: ltm_vertical_profiles}. It is evident that, there exists a substantial disparity between observed and simulated profiles without aerosols, both in profile shape and mean temperatures ($\sim$  2--5°C). Incorporating aerosols into the simulations largely mitigates these temperature differences, resulting in simulated vertical profiles resembling LTM as observed in this field study. Also, the impact of including aerosols in radiative processes extends several hundred meters into the boundary layer. However, still a significant offset in temperature (more than 2$^o$C) occurs above 20 m, this might be due to large-scale temperature advection (see supplementary figure S2) and day-to-day variations as discussed below.

In Figure \ref{fig: ltm_vertical_profiles}, we present two types of averaging both for simulations and field observations. These plots elucidate two types of variabilities in the temperature profile. We have considered observations and simulated data at five-minutes time interval, starting from local sunset time to the next four hours on each of the eighty days. For this part of the discussion, concentrate on observed profiles (green color) and simulated profiles with aerosols (blue color). The mean temperature profiles in Figures \ref{fig: ltm_vertical_profiles}a and \ref{fig: ltm_vertical_profiles}b, are represented by solid green-lines (observation) and dashed blue-lines for simulations. Day-to-Day variability in the temperature profiles for eighty days is depicted in Figure \ref{fig: ltm_vertical_profiles}a. For this plot, temperature measurements for each day within the 4-hour window following local sunset are averaged to obtain a single mean temperature profile for that day. This process is repeated for eighty-days in the dataset, resulting in a collection of eighty-mean temperature profiles. Now the average of these eighty profiles is the solid green-line (for observation) and dashed blue-line (for simulation with aerosols) in Figure \ref{fig: ltm_vertical_profiles}a. Green-shaded region is obtained by calculating variability ($\pm 1 \sigma$) of observed temperature at each height from the mean value of observation at that height. Shaded region indicates day-to-day variability of temperature in observation, resulting from the prevailing atmospheric conditions. Similarly blue-shaded region indicates day-to-day variability of simulated temperature profile for given input temperature initialization and observed boundary condition.   

We can also investigate temporal variability of the temperature profiles (see Figure \ref{fig: ltm_vertical_profiles}b). For this analysis, we consider five-minute interval separated 48 (4 hr x 12/hr) temperature profiles on each day for eighty-days. Here, we take averages across eighty-days for corresponding time profiles, we get a set of 48-mean temperature profiles. The average of these 48-profile is same as the one we got previously and is plotted as solid green-line (observation) and dashed blue-line (for simulation with aerosols). As above, for the set of 48-temperature profiles calculating the variability ($\pm 1 \sigma$) of temperature at each height from the mean value at that height, we have plotted the shaded green (observation) region and shaded blue (for simulation with aerosols) region. It is evident that, in the surface layer up to 20--30 m, the day-to-day variability (in Figure \ref{fig: ltm_vertical_profiles}a) is many times greater than the temporal variability (in Figure \ref{fig: ltm_vertical_profiles}b). The result indicates that, time to establish observed temperature profile on a given day is very short, and most of the variation originates from the day-to-day variation in the atmospheric conditions including temperature, water vapor in the atmosphere, diurnal history of solar insolation and probably even the aerosol distribution.

\begin{figure}[ht]
	\centerline{\includegraphics[width=\textwidth]{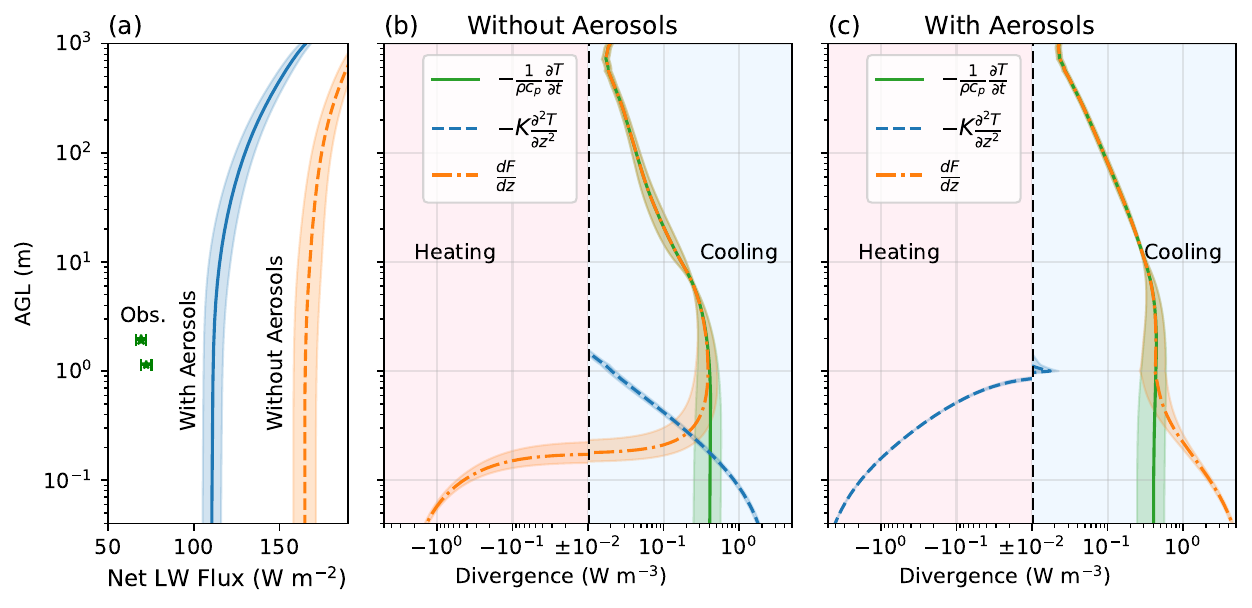}}
	\caption{Mean radiative flux and its divergence with and without aerosols from local sunset to next 4 hours over the daily mean profiles of 80 days analysis. (a) Net LW radiative flux with and without aerosols from simulations. Flux measurements from the field experiments are also marked. (b) Vertical variation of radiation, conduction, and net divergence without aerosols and (c) with aerosols. Here, absolute divergence less than 0.01 W m$^{-3}$ has been shown as $\pm10^{-2}$ W m$^{-3}$.  Spread at any height is one standard deviation over 4 hours from sunset (shading around line plot).}
	\label{fig: radiative flux evolution}
\end{figure}

Scattering, absorption, and emission of radiation from aerosols change the net radiative flux and net radiation divergence, leading to different temperature profiles in the atmosphere \citep{liou2002introduction}. 4-hour (from sunset) mean net radiative flux with/without aerosols, as well as net LW flux obtained from radiation sensors, have been presented in Figure \ref{fig: radiative flux evolution}a (green marker). The shaded region shows one standard deviation spread over 4 hours. The maximum spread in both simulations and observations is less than 6 W m$^{-2}$ at 1.14 m AGL, which signifies that simulations and observations follow similar variability over the evolution. Downward LW fluxes obtained from both radiation sensors show an offset of 4 W m$^{-2}$, a possible relative uncertainty in radiative flux measurement. However, net radiative flux from simulations is ~100 W m$^{-2}$ higher than the observed one if aerosols are not accounted for, and the above difference reduces to 40--50 W m$^{-2}$ if aerosols are accounted for in the model. These offsets in downward LW flux can be further reduced if other major greenhouse gases like Carbon Dioxide (CO$_2$) and Ozone (O$_3$) are accounted for in the radiation model.

Radiation, conduction, and net divergence  with/without aerosols are shown in Figure \ref{fig: radiative flux evolution}b, c (Equation \ref{eq: 1_dim_cond_radiation_model}). When aerosols are not accounted for in the model, radiation divergence induces heating within a few decimeters above the surface, but conduction divergence causes cooling. In this region, cooling induced by conduction divergence dominates over warming induced by radiation divergence and produces net cooling close to the surface. After a few decimeters above the surface, conduction divergence weakens, and radiation divergence dominates over the remaining column of the atmosphere. Overall, in the absence of aerosols, radiation and conduction divergence together lead to net divergence (cooling) of the whole column, but net divergence decreases monotonically with height and does not cause any preferential cooling, which is required for the development of LTM.    

In the presence of aerosols, radiation divergence is substantial near the surface and dominates over conduction divergence over the whole column (Figure \ref{fig: radiative flux evolution}c). However, as heating induced by conduction divergence decreases rapidly away from the surface and cooling caused by radiation divergence dominates (Equation \ref{eq: 1_dim_cond_radiation_model}). It leads to locally enhanced net divergence between 0.5 to 2 m above the surface and makes the net divergence profile non-monotonous with height. This enhanced net divergence causes preferential cooling, which leads to LTM development. Interplay of these flux divergence terms, at the surface, is further complicated by the presence of penetrative convection system  (see \citet{kaushal2024penetrative}) driven by radiative cooling. We also observe that aerosol-induced radiation divergence (net cooling) after a few decimeters from the surface is higher than the no-aerosols conditions, extending for a few hundred meters. Although the net divergence (cooling) decreases with height, it can induce significant temperature change over the night in ABL.

%%%%%%%%%%%%%%%%%%%%%%%%%%%%%%%%%%%%%%%%%%%%%%%%%%%%%%%%%%%

\begin{figure}[ht]
	\centerline{\includegraphics[width=\textwidth]{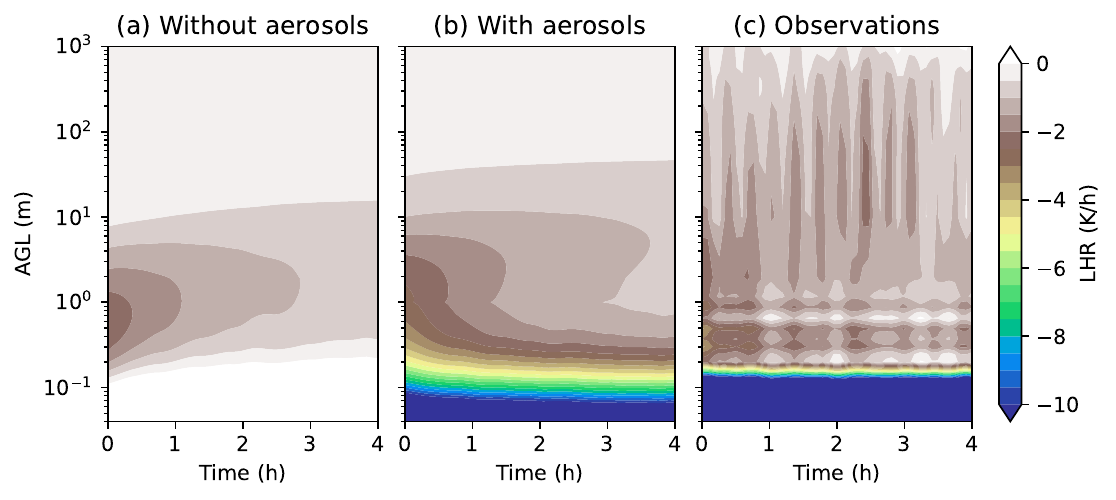}}
	\caption{Temporal variation in mean LHR from local sunset to next 4 hours up to 1 km (a) Without aerosols: when aerosol-radiation interactions are not accounted for in the model (b) With aerosols: When aerosol-radiation interactions are accounted for, (c) Observations; derived LHR from observed temperature profiles using Equation \ref{eq: 1_dim_cond_radiation_model}. Derived LHR from observations are in better agreement with aerosol-accounted LHR, compared to without-aerosol LHR.}
	\label{fig: LHR}
\end{figure}

\subsection{Longwave heating rate (LHR) after sunset}
Temporal variation in day-to-day mean LHR over 80 days from local sunset to the next 4 hours are shown in Figure \ref{fig: LHR}. Since the net heating rate and conduction heating rate can be directly estimated from the temperature profile, LHR profiles have been derived using Equation (\ref{eq: 1_dim_cond_radiation_model}) under the assumption of negligible horizontal advection and vertical mixing. When aerosols are not accounted for, the mean LHR near the surface is weaker compared to the LHR from the field observations and aerosol-accounted simulations (Figure \ref{fig: LHR}). Simulations without aerosols show positive LHR and, hence, radiative warming within a few centimeters from the surface.  Above the warming region, radiative cooling of $\approx$2 K/h is observed near the ground. A comparable radiative cooling rate has also been reported without aerosols in \citet{ha2003radiative} and  \citet{steeneveld2006modeling}. In contrast, observations and aerosols-accounted models show intense cooling within 1 m AGL where LHR goes less than -10 K/h. Similar values of LHR of several K/h during sunset have also been reported by \citet{steeneveld2010observations}, but all LW radiation models systematically underestimate radiative cooling by one order of magnitude, which could be attributed to the absence of aerosols-radiation interaction in the models. Further, when we don't account for aerosols in the models, it can lead to an offset of $\approx 4$ K in the model (Figure \ref{fig: ltm_vertical_profiles}), which can affect the onset, growth, and intensity of fog.  Here, we show that LHR profiles from simulations and field experiments are of the same order only if aerosols-radiation interactions are allowed in the model.  

The temporal and vertical evolution of LHR profiles from aerosol-accounted simulations is in good agreement with derived profiles from field observations (Figure \ref{fig: LHR}). During local sunset, radiative cooling of more than 1 K h$^{-1}$ is observed up to 10 m AGL in aerosol-accounted models as well as in field observations, which causes intense cooling close to the ground. Further, LHR decreases sharply with height, but a weak negative LHR can be observed till a few hundred meters. We believe that LHR fluctuations in derived LHR are due to the advection of temperature and humidity profiles, which has not been accounted for in the present model. Although LHR spread in observations is large compared to simulations, mean LHR from aerosol-accounted simulations is in better agreement with the observed mean LHR compared to simulations. Moreover, it is clear from derived LHR and aerosol-accounted simulation that LHR can be lower than -5 K h$^{-1}$ very close to the ground which might play an important role in land-atmosphere coupling.

\subsection{Radiation divergence and LTM in the presence of fog/cloud} \label{sec: fog_thickness}

Based on the past field observations, it has been speculated that LTM disappears if cloud passes over it \citep{ramdas1932vertical, oke1970temperature, mukund2014field, blay2015lifted}. We also observe that LTM disappears in most of the fog events but sustains in one fog event (Figure \ref{fig: LTM observations}, on 18$^{th}$ December). Hence, to investigate the role of different thicknesses as well as different base heights of fog on LTM, we have considered four idealized layers of fog/cloud in simulations. These layers are located between; surface to 10 m (T1), surface to 90 m (T2), 10 m--100 m (T3), and 300 m--390 m (T4). Although T1 and T2 both represent fog touching the ground, T1 is shallower than T2. T3 and T4 both represent fog/cloud whose thickness is the same, but its base height is different. Further, the size distribution and concentration of fog are not site-specific and are taken from the OPAC database \cite{hess1998optical}. For the sake of simplicity, the concentration and size distribution of fog do not change within a fog layer and show sharp changes across the boundary. In this set of  analysis, aerosol profile considered is same as earlier (Figure \ref{fig: aerosol concentration}), in addition, fog droplets were considered with a modified gamma distribution having a total concentration of 15 cm$^{-3}$. 

For each fog thickness, we have simulated the growth and dissipation of LTM profiles, radiative flux, and its divergence in the presence of the fog layer. In all three simulations, the initial 2 hours of the simulations are run without the fog layer to allow the LTM to develop. After 2 hours of simulations, the fog layer is activated, where the evolution of different parameters like temperature, radiative flux, and divergence under the combined effect of aerosols and fog are observed for the next 2 hours of simulation.

\begin{figure}[ht]
	\centerline{\includegraphics[width=0.8\textwidth]{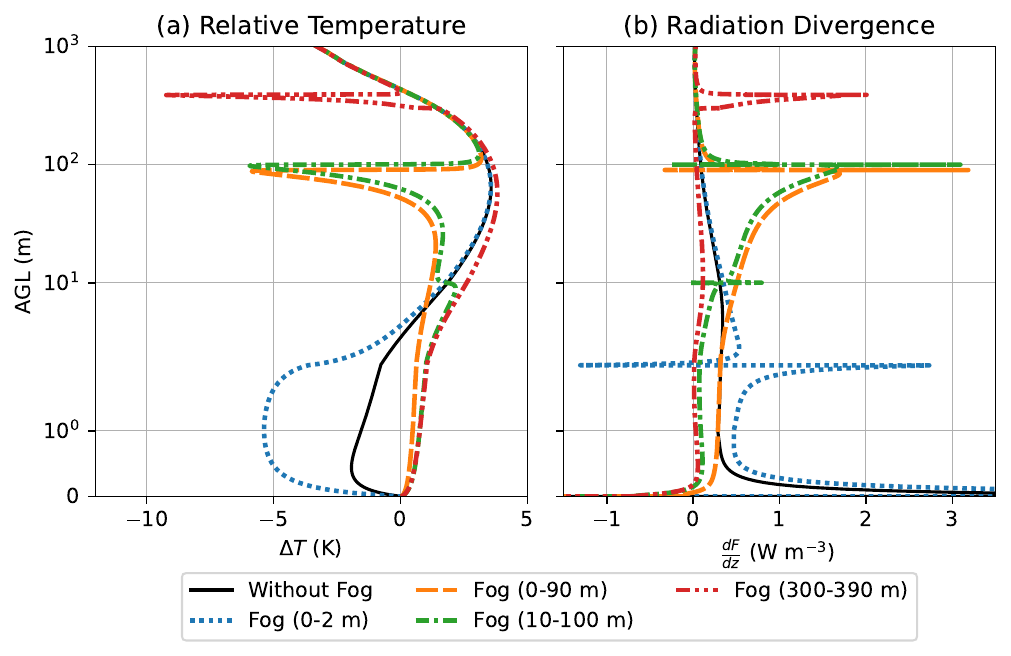}}
	\caption{Vertical profiles of (a) relative temperature and (b) radiative divergence in the presence of different fog layers.}
	\label{fig: temp_rad_div_in_fog}
\end{figure}

 Figure \ref{fig: temp_rad_div_in_fog}(a) shows relative temperature profiles compared to the surface temperature in the presence of different fog/cloud layers as well as without it. Without any fog/cloud, we observe LTM profiles of 2 K, usually observed in our field experiments. However, LTM intensity and height increase in the presence of shallow fog of 2 m thickness (Case T1) due to enhanced radiation divergence near the ground (Figure \ref{fig: temp_rad_div_in_fog}(b)). It is due to an increase in radiation interaction with fog droplets and clear skies. A similar case, like maintenance of LTM in fog has been observed during a fog event (Figure \ref{fig: LTM observations}). At the fog top and bottom, we notice a sharp change in temperature profiles and radiative divergence, which is due to idealized boundary conditions, i.e., the sharp change in fog concentration at the boundary. After the fog top, temperature and radiative divergence tend towards without fog condition. With an increase in fog thickness (Case T2), the optical thickness of fog becomes so large that it behaves like an opaque sheet for the air layer near the ground. Hence, radiative flux evolution becomes practically independent above the fog layer and near the ground, and LTM does not appear. We have observed an increase in downward LW radiation in the presence of fog in the simulations (not shown here), which has also been observed in the field experiments (Figure \ref{fig: LTM observations}). Hence, we observe an inversion profile rather than an LTM profile. Further, with an increase in the cloud base height, keeping the same thickness (Case T3 and T4), we observe that radiation divergence weakens near the ground, which results into an inversion profile of higher intensity. 

We observe a significant temperature drop at the fog top. Many field observations and simulations have reported intense fog top cooling \cite{roach1976physics, nishikawa2004radiative, koravcin2014marine, yang2020impact}, but we note here that the present model is one-dimensional and does not account for gravitational settling, downward draft, and mixing due to turbulent convection (because of negative buoyancy). With such limitations, temperature drop within the fog layer or fog top/bottom indicated here can be significant. Further, vertical mixing induced by fog top cooling might lead to temperature convergence within the fog layer \cite{price2011radiation}. Moreover, the energy released during phase change in foggy conditions offsets the cooling caused by radiation divergence. As fog top height increases, cooling caused by radiation divergence extends across the fog layer, which results in reduced observable cooling within the fog layer. If fog top height is substantial and fog duration is brief, temperature drop because of fog top cooling is not significantly noticeable (See supplementary figure S3). However, our interest here is to test the radiative effect of the fog layer, especially the impact of fog on radiation divergence and the sustenance of LTM. Therefore, fog micro-physics and related dynamics will not be discussed here. We observe that enhanced radiation divergence at the fog top is responsible for the intense cooling at the fog top.

\section{Discussions} \label{sec: discussion}
LTM characteristics vary significantly across different field experiments \cite{oke1970temperature, mukund2014field, blay2015lifted}. In the present field experiments, the observed mean LTM height is close to the height observed by \citet{mukund2014field}, whereas the observed intensity is lower. \citet{mukund2014field}, who studied LTM at one location with modified surface properties, have shown that LTM height does not depend significantly on the surface type. We expect that it strongly depends on other parameters, especially the aerosol concentration profiles, which have high spatial and temporal variability \cite{wu2021urban}. An extensive field experiment with simultaneous profiling of aerosols and LTM would unravel the direct correlation between aerosol vertical distribution and variation in LTM height as reported in field observations \cite{oke1970temperature, mukund2014field, blay2015lifted}. 

The results obtained from the one-dimensional conduction-radiation model, which incorporates radiation interactions with a representative aerosol profile, closely agree with the observed temperature profiles after the evening transitions. This agreement also extends to predicted and measured radiation divergence during the evening transition in the present study and other field observations \cite{steeneveld2010observations}. However, considerable variability in aerosol vertical distribution, particle concentrations, sizes, chemical composition, and their intricate interactions with radiation is reported by \cite{forster2021earth}. In this survey, the total aerosol concentration at a few sites was found to exceed $10^{5}$ cm$^{-3}$, but most of these observations have been performed a few meters away from the ground, and aerosols within 1--2 m above surface have not been investigated. However, aerosol concentration within 1 m can be 10-100 times higher than the concentration measured a few meters above the ground \cite{devara1995real, chate2004field}.

Overall, in light of significant uncertainties associated with spatial and temporal variability of aerosol particle concentrations and properties, we use aerosol concentration, profile, and properties presented in Section \ref{sec: aerosol_con_and_prof}. However, detailed profiling of aerosols at the observation site might further reduce the offset in LTM height and offset in temperature profiles at different vertical levels away from the surface. Despite uncertainties in the aerosol profile, we demonstrate that LTM simulations appear in good agreement with the field observations only when aerosols are included. It is also consistent with laboratory-scale LTM observations by \citet{mukund2014field} that LTM intensity decreases with a decrease in aerosol concentration in the test section. Moreover, the evolution of radiative flux, divergence, and temperature are in better agreement with field observations if aerosols are accounted for in the model. Therefore, results presented here clearly emphasise the  need to include aerosol-induced radiative cooling in current radiation models used in stable, nocturnal ABL.

The model employed in this study is a one-dimensional conduction-radiation model that considers aerosols. However, it does not incorporate horizontal advection, vertical convection/mixing, or other dynamical processes. Hence, the model has limitations in accounting for the changes in the temperature profile due to horizontal advection and vertical mixing. Although we show from current field observations that radiation divergence plays a crucial role in LTM development and its decay, a better comparison with simulations and field observations can be achieved if other dynamical processes like vertical mixing in unstable layer near the ground are also accounted for in the model \citep{mahrt2014stably}. 

A basic calculation shows that a 0.5 W m$^{-3}$ radiation divergence can lead to a cooling rate of 1.8 K/h, which can influence many meteorological phenomena after sunset like near-surface temperature inversion, fog, pollution dispersal, drainage flow, and nocturnal jet in the NBL. Since occurrence, dissipation, and intensity of fog and drainage flow are sensitive to small temperature changes, prediction of these phenomena can be improved if aerosols-radiation interaction in the LW region is accounted for in the current forecasting model.

\section{Conclusions} \label{sec: conclusion}
%%% working here

During evening transition and later in the night, radiative cooling strongly modulates the thermal structure of the ABL near ground which has significant implications in micro-meteorology and agriculture. Most of the radiation models are not able to produce it satisfactorily \cite{steeneveld2010observations, steeneveld2014current} which might be due to missing aerosol-radiation interaction in LW region \citep{zdunkowski1976one, coantic1971interaction, andre1982nocturnal, mukund2010hyper}.
In this paper, we have presented results that elucidate the role of radiative cooling due to aerosols on the thermal structure of the ABL during and after evening transitions through extensive field observations and numerical simulations. From our field experiments, we have demonstrated that the Lifted Temperature Minimum (LTM) generally occurs in the nocturnal boundary layer under calm and clear sky conditions, typically appearing during the evening transition and intermittently disappearing during the night. The persistence of LTM depends on specific conditions and it is not observed during the daytime when solar heating and convection dominate.

Our field observations reveal that LTM occurrence are strongly influenced by factors like mean wind speed, turbulent kinetic energy, downward and upward longwave fluxes, as well as sensible heat flux. The probability of occurrence of LTM occurrence increases with a decrease in the above parameters. Notably, LTM's disappearance with increasing wind speed suggests that near-surface advection is not the primary cause.  Further, LTM is observed in both seasons (winter and spring), and there is no significant change in its characteristics across the seasons. Observed LTM intensity and height are 2.2$\pm$0.6 K and 0.31$\pm$0.11 m (mean and one standard deviation), respectively. These values are in a similar range as reported by \citet{mukund2014field} and \citet{blay2015lifted}. 

Simulations using a one-dimensional, conduction-radiation model that accounts for aerosols show that LTM cannot form in the absence of aerosols, even under favorable conditions. Without aerosols, the net divergence of longwave radiation decreases monotonically with height, leading to a typical temperature inversion profile. However, the presence of aerosols results in a non-monotonic net divergence, causing preferential cooling near the ground and leading to LTM development. Simulated LTM height and intensity match with field observations when aerosols are considered. Further analysis with idealized fog layer simulations reveals that the presence of fog modulates downward longwave flux and radiation divergence depending upon the fog base height and thickness. In case of shallow fog near the ground, LTM strengthens with an increase in radiation divergence. Both radiation divergence and LTM intensity decrease with an increase in fog thickness, and if the fog becomes optically thick, LTM disappears completely, and we observe an inversion profile. This behavior aligns with observations in the field, supporting the model's microphysics of radiative transfer.

We have also investigated the reasons for the underestimation of longwave heating rate (LHR) by radiation models during evening transitions, as reported by \citet{steeneveld2010observations}. Our findings, along with simulations with and without aerosols, demonstrate that LHR from simulations and field experiments better agree when aerosols are considered in the model. Aerosol-induced LHR is not limited to LTM occurrence during evening transitions; its effects extend several hundred meters above ground level and can influence various meteorological phenomena such as the development of the nocturnal boundary layer, temperature inversion, mist, fog, and pollution dispersion, ultimately affecting the stability of the boundary layer. Incorporating aerosol-radiation interactions in longwave radiation models will lead to improved forecasts of these phenomena.

%% limitations and future scope 

 Although we have shown the role of aerosols in radiation divergence as well as LTM occurrence in calm and clear sky conditions from 1-dimensional model, its parameterized version needs to be developed and tested in NWP models. Further, current study does not account for dynamical modeling which can be important near the ground. Although penetrative convection formed near the ground due to radiative cooling  has been studied by \citet{kaushal2024penetrative}, LTM maintenance against convective instability and turbulent flux divergence in the unstable layer of LTM needs to be investigated in future. Other LTM characteristics like its transient behavior, appearing and disappearing nature during night are currently under investigation from observations and simulations.

%%%%%%%%%%%%%%%%%%%%%%%%%%%%%%%%%%%%%%%%%%%%%%%%%%%%%
\section{Acknowledgements}\label{acknowledgements}
We thank the Department of Science and Technology, Government of India, for co-funding this project through the Technical Research Centre (TRC) program at JNCASR, Bengaluru, India. Additionally, we extend our thanks to Bangalore International Airport Limited (BIAL), Bengaluru, India, for co-funding this project and for granting access to the runway area at Kempegowda International Airport Bengaluru (KIAB) to establish our observation site, as well as providing other logistical support. Furthermore, we acknowledge the National Supercomputing Mission (NSM) program at JNCASR for providing the computational facility necessary for this research project.

\section{Conflict of interest}
We herewith declare that we do not have any conflict of interest in the work reported in this paper.

\selectlanguage{english}

%\bibliography{common_references}

%% I will add supplementary material with updated page number in the last of this main pdf.

% \pagebreak
% \include{revised_manuscript_1/Supplementary information for Investigation of the Thermal Structure in the Atmospheric/supp}

\end{document}